\RequirePackage{fix-cm}
\documentclass[smallextended]{svjour3}       % onecolumn (second format)
\smartqed  % flush right qed marks, e.g. at end of proof
\usepackage{graphicx}
\usepackage{braket}
\usepackage{amsfonts}
\usepackage{amsmath}
\usepackage{amssymb}
\usepackage{enumerate}

\usepackage{hyperref}
\usepackage{xcolor,soul}
\usepackage{wasysym}
\usepackage{times}
\usepackage{mathrsfs}

\hypersetup{
pdfstartview={FitH},% fits the width of the page to the window
pdftitle={2018},    % title
pdfauthor={Luis E. F. Foa Torres},  % author
colorlinks=true,    % false: boxed links; true: colored links
linkcolor=blue, % color of internal links
citecolor=blue,   % color of links to bibliography
filecolor=blue, % color of file links
urlcolor=blue   % color of external links
}

\newcommand{\bk}{{\bf k}}

\begin{document}

\title{Topological states of non-Hermitian systems}

\titlerunning{Topological states of non-Hermitian systems}        % if too long for running head

\author{V. M. Martinez Alvarez \and
        J. E. Barrios Vargas  \and M. Berdakin \and L. E. F. Foa Torres %etc.
}


\institute{V. M. Martinez Alvarez \at
 			  Departamento de F\'{\i}sica, Laborat\'orio de F\'{\i}sica Te\'orica e Computacional, Universidade Federal de Pernambuco, Recife 50670-901, PE, Brazil
%              \email{fauthor@example.com}           %  \\
           \and
           J. E. Barrios Vargas \at
 			  Departamento de F\'{\i}sica, Facultad de Ciencias F\'{\i}sicas y Matem\'aticas, Universidad de Chile, Santiago, Chile
%              Tel.: +123-45-678910\\
%              Fax: +123-45-678910\\
%              \email{fauthor@example.com}           %  \\
           \and
           M. Berdakin \at
 			  Departamento de F\'{\i}sica, Facultad de Ciencias F\'{\i}sicas y Matem\'aticas, Universidad de Chile, Santiago, Chile
%              Tel.: +123-45-678910\\
%              Fax: +123-45-678910\\
%              \email{fauthor@example.com}           %  \\
           \and
           L. E. F. Foa Torres \at
 			  Departamento de F\'{\i}sica, Facultad de Ciencias F\'{\i}sicas y Matem\'aticas, Universidad de Chile, Santiago, Chile,   
%              Tel.: +123-45-678910\\
%              Fax: +123-45-678910\\
              \email{luis.foatorres@uchile.cl}           %  \\
}

\date{Received: 11 May 2018 }%/ Accepted: date

\maketitle

\begin{abstract}
Recently, the search for topological states of matter has turned to non-Hermitian systems, which exhibit  a rich variety of unique properties without Hermitian counterparts. Lattices modeled through non-Hermitian Hamiltonians appear in the context of photonic systems, where one needs to account for gain and loss, circuits of resonators, and also when modeling the lifetime due to interactions in condensed matter systems. Here we provide a brief overview of this rapidly growing subject, the search for topological states and a bulk-boundary correspondence in non-Hermitian systems.  

\end{abstract}

\section{Introduction}
\label{intro}
Over the last three decades, a series of breakthroughs have completely reshaped our understanding of condensed matter physics creating a new research field, topological states of matter~\cite{hasan_colloquium_2010}. Starting with the discovery of the integer quantum Hall effect~\cite{von_klitzing_new_1980}, which was closely followed by arguments affirming its topological origin~\cite{thouless_quantized_1982} and the possibility of obtaining it even without Landau levels~\cite{haldane_model_1988}, captivating insights~\cite{kane_quantum_2005,bernevig_quantum_2006} led to the rapid discovery of topological insulators in two~\cite{konig_quantum_2007} and three dimensions~\cite{hsieh_topological_2008}.
Within this young field, the treasure hunt is sprouting into fascinating new directions including Weyl semimetals~\cite{vafek_dirac_2014} (gapless but topological phases), topological states in  driven (out of equilibrium) systems~\cite{oka_photovoltaic_2009,lindner_floquet_2011,foa_torres_multiterminal_2014} and non-Hermitian lattices~\cite{rudner_topological_2009,diehl_topology_2011,yuce_topological_2015,san-jose_majorana_2016,lee_anomalous_2016,leykam_edge_2017}.

In this brief review article we focus on the emerging research front devoted to the search for topological states in non-Hermitian systems, specifically edge or boundary states appearing on lattices represented by a non-Hermitian Hamiltonian. Writing a review-style article of such a young and rapidly growing field is challenging and we hope, in spite of not being fully comprehensive, that it helps readers at the crossroads between the different converging communities. A very first question is what lies behind the term non-Hermitian? Although quantum mechanics is traditionally formulated in terms of Hermitian operators (so that their eigenvalues are warranted to be real), non-Hermitian Hamiltonians~\cite{rotter_non-hermitian_2009,moiseyev_non-hermitian_2011} have gained a physically relevant space from two different viewpoints. The first is more fundamental and touches mathematical physics, it has to do with ${\cal PT}$-invariant alternatives to Hermitian quantum  mechanics~\cite{Bender_prl_1998}.
One says that a Hamiltonian is ${\cal PT}$-invariant when it commutes with the composed parity-time ${\cal PT}$ operator. Interestingly, in many cases the Hermiticity requirement (for the eigenvalues to be real) can be lessened and requiring ${\cal PT}$ invariance of the Hamiltonian suffices (if the states happen to be ${\cal PT}$ symmetric then their eigenvalues are real in the ${\cal PT}$-unbroken phase). In the second viewpoint, which will be our focus throughout this article, non-Hermitian Hamiltonians are regarded as \textit{effective} Hamiltonians (where the non-Hermitian part serves different purposes). Examples include open quantum systems~\cite{rotter_non-hermitian_2009} \footnote{For example, the use of non-Hermitian Hamiltonians in the field of quantum transport is well known. Typically one considers a finite sample connected to electrodes and, after tracing over the degrees of freedom corresponding to the infinite electrodes one obtains an effective non-Hermitian Hamiltonian for the sample.}, systems with gain and loss (as found in photonics~\cite{ruter_observation_2010,longhi_PT_2017,ozawa_topological_2018}) or systems where the non-Hermiticity models the finite lifetime introduced by electron-electron or electron-phonon interactions~\cite{Kozii2017,yoshida_non-hermitian_2018}. Also, the effects of non-Hermiticity has been explored in the context of localization-delocalization transitions following the pioneering works by Hatano and Nelson~\cite{Hatano_1996,Hatano_1997,Hatano_1998}, biological systems~\cite{Nelson_Non_Hermitian_1998,Amir_2016,murugan2017topologically}, Weyl semimetals~\cite{Xu_Weyl_2017,zyuzin2018flat,Cerjan_2018,Molina_2017,Molina_eng_EPs_2018}, as well as the interplay between topology and dissipation~\cite{diehl_topology_2011,bardyn2013,budich2015topology,budich2015dissipative}.

In the following we address a selection of relevant issues contextualizing recent publications. Section~\ref{sec:non-hermiticity} introduces a few key fingerprints of non-Hermitian systems including, notably, the so-called exceptional points. Section~\ref{sec:bulk-boundary} provides an overview of the challenges and current debate around the bulk-boundary correspondence in non-Hermitian lattices. Section~\ref{sec:classification} addresses the attempts to classify the topological phases. Section~\ref{sec:experiments} focuses on recent experiments in this field and, finally, Section~\ref{sec:conclusions} provides our final remarks.

\section{Unique features imposed by non-Hermiticity}
\label{sec:non-hermiticity}

As mentioned before, we will consider regular lattices described by non-Hermitian Hamiltonians~\footnote{One must notice that most of the literature on non-Hermitian Hamiltonians is not focused on lattices as common in Solid State Physics. The study of this type of systems has been ignited more recently motivated by the search of topological states.}. This effective description stems because either in the particular system one needs to introduce gains and losses to describe observations~\footnote{In photonics, for example, gain and loss occur quite naturally. Losses due to absorption are almost inescapable while gain media are crucial in lasers~\cite{Silfvast_lasers_2004}.}, because the system is open to other degrees of freedom on which we are not interested in, or because of interactions providing a finite lifetime to the quasi-particle excitations. In this sense, the non-Hermitian part of the Hamiltonian can be considered as resulting from the effect of an environment on a subspace of the full Hilbert space~\cite{rotter_non-hermitian_2009}. From this viewpoint, one might imagine that there is a parent Hermitian system from which everything could be solved. Why one might then insist on dealing with the effective non-Hermitian Hamiltonian? In many cases such parent Hamiltonian may not be available or it is just too complex, thereby motivating the use of such effective description. In the case where the non-Hermiticity encodes the effect of interactions, this seems the only simple way to keep a single-particle picture while capturing the most important ingredients.

What are the key differences between Hermitian and non-Hermitian systems and when do they manifest? A first observation is that not all non-Hermitian Hamiltonians show properties truly unique to the non-Hermiticity. For example, adding an imaginary constant times the identity operator to an Hermitian Hamiltonian adds a trivial non-Hermiticity, in the sense that the eigenvectors remain unchanged cause the two terms commute. On the other hand, sometimes a non-Hermitian Hamiltonian with real eigenvalues could be \textit{Hermitian in disguise} (also called \textit{crypto-Hermitian})~\cite{Smilga2009,feinberg_1999}. In such case there is a similarity transformation which takes the Hamiltonian to a Hermitian form~\cite{Smilga2009,feinberg_1999}.

One of the unique non-Hermitian features is the existence of singular points where not only the eigen-energies but also the eigenfunctions coalesce. At these singularities, that Kato~\cite{kato_1966} called \textit{exceptional points}, the Hamiltonian becomes \textit{defective}, this is, it lacks a full basis of eigenvectors~\cite{heiss_physics_2012} (and it cannot be taken to a diagonal form) .
  
Take as an example the following matrix:
\begin{equation}
{\cal H} = 
 \begin{pmatrix}
  1 & 1  \\
  0 & 1  \\
 \end{pmatrix},
 \end{equation}
which is a $2\times 2$ matrix representing a Jordan block. This matrix has the eigenvalue $\lambda=1$ which is a double root of the characteristic polynomial, but since there is a single eigenvector associated to it, the matrix is defective. In this case one says that algebraic multiplicity of the eigenvalue $\lambda=1$ is $2$ (the number of times $\lambda=1$ is a solution of the characteristic polynomial) while the geometric multiplicity is $1$ (the dimension of the space spanned by the eigenvectors).

Generically, one may consider a Hamiltonian of the form ${\cal H}={\cal H}_0+\lambda {\cal H}_1$, where the parameter $\lambda$ quantifies the strength of the interaction between ${\cal H}_0$ and ${\cal H}_1$ (both assumed Hermitian). For real $\lambda$, interacting levels repel each other and do not cross as $\lambda$ changes~\cite{vonNeuman1929}. But the same levels do coalesce if $\lambda$ is taken to the complex plane. Typically, the coalescence is a square root singularity of the spectrum as a function of $\lambda$, an exceptional point. The importance of these points was remarked by Berry~\cite{berry_pancharatnam_1994,berry_geometric_2010} in connection with an earlier discovery by Pancharatnam~\footnote{As Berry elegantly puts it ``At the heart of Pancharatnam’s discovery of a remarkable phenomenon in light propagation in absorbing crystals is the behavior of eigenstates at a nonhermitian degeneracy" (see this \href{http://indico.ictp.it/event/a0368/session/9/contribution/6/material/0/0.pdf}{url}).}. 
% phrase taken from http://indico.ictp.it/event/a0368/session/9/contribution/6/material/0/0.pdf

Even for the simplest $2\times 2$ Hamiltonian, there are several peculiar phenomena associated to the exceptional points including~\cite{heiss_physics_2012}: 
\begin{enumerate}[(a)]%for small alpha-characters within brackets.
\item At a finite distance from the EP one has two linearly independent eigenvectors $\ket{\psi_1}$ and $\ket{\psi_2}$ (their left counterparts being $\bra{\phi_1}$ and $\bra{\phi_2}$ and normalized so that $\braket{\psi_j | \phi_j}=1$), while at the EP there is only one and $\braket{\psi_j | \phi_j}$ vanishes; 

\item The spectrum depends strongly on $\lambda$ in the vicinity of an EP, the derivative with respect to $\lambda$ of the eigenvectors and eigenvalues diverges at an EP;

\item Repeated encircling of an EP generates the pattern $\ket{\psi_1}\rightarrow -\ket{\psi_2} \rightarrow -\ket{\psi_1} \rightarrow \ket{\psi_2} \rightarrow \ket{\psi_1}$ when moving counterclockwise, thus having a $4\pi$ periodicity. Doing the same clockwise gives a different sign, thereby showing a chiral behavior~\cite{heiss_chirality_2001}\footnote{The change of a state when encircling an EP, also called \textit{chiral state conversion}, establishes the ground for the so-called topological properties of the exceptional points. One must, however, distinguish between the name topological used in that context and that used in the context of topological states of matter and topological insulators. We will come to the latter point in the next section.}. Interestingly, a recent study predicts that this chiral state conversion can take place even without encircling the EP~\cite{Hassan2017} (as long as the loop is in its vicinity~\cite{Hassan2017b}).
\end{enumerate}

These striking properties have been observed experimentally in microwave experiments~\cite{dembowski2001,stehmann2004,Lee_observation_2009,hu2017exceptional}, nuclear magnetic resonance~\cite{alvarez2006}, optical ~\cite{ghosh2016exceptional} and microwave waveguides~\cite{doppler2016dynamically}, and in optomechanical setups ~\cite{xu2016topological}. Other experiments in ferromagnetic waveguides threaded by a magnetic field allow tuning the dynamic pathway around EPs to explore the chiral conversion effect~\cite{zhang2018dynamically}. The authors~\cite{zhang2018dynamically} show that the chiral conversion obtained when encircling an EP is turned on (off) when the starting/end point of the dynamic belongs to the unbroken (broken) ${\cal PT}$ phase space.

Beyond the $2\times 2$ case, EPs can also be of higher-order~\cite{heiss_physics_2012}. Most of the available literature focuses on the coalescence of a few levels, typically $3$~\cite{ryu2012,Eleuch2016,Jing2017,Pick17,Hodaei2017}, but one also have coalescences of very high-order~\cite{Graefe2008}, even scaling with the matrix size~\cite{Martinez_Non-Hermitian_2018} or condensing, like an Aleph~\footnote{In the brief story \href{https://en.wikipedia.org/wiki/The_Aleph_(short_story)}{`The Aleph'}, written by Jorge Luis Borges in 1945, an Aleph `is one of the points in space that contains all other points'.}, the full spectrum on one point. Less is known about these latter, with recent results pointing to a highly unusual situation where all the states could be localized~\cite{Martinez_Non-Hermitian_2018}. %\VM{even those in the bulk}.

We emphasize once more that, although exceptional points occur in a region of zero measure in parameter space and the coalescence is usually fragile against perturbations, their consequences do extend to their vicinity and can be measured in experiments. One way of quantifying these effects around EPs is through the \textit{phase rigidity}. The right and left eigenstates of ${\cal H}$ obey the relations: ${\cal H} \ket{\psi_{\alpha}} = \varepsilon_{\alpha} \ket{\psi_{\alpha}}$ and $\bra{\phi_{\alpha}} {\cal H}= \varepsilon_{\alpha} \bra{\phi_{\alpha}}$. When ${\cal H}$ is non-Hermitian, $\bra{\phi_{\alpha}} \neq \bra{\psi_{\alpha}}$. A measure of the eigenfunctions' biorthogonality is the phase rigidity $r_{\alpha}$ defined as~\cite{rotter_non-hermitian_2009}: 
\begin{equation}
r_{\alpha}= \frac{\braket{\phi_{\alpha} | \psi_{\alpha}}}{\braket{\psi_{\alpha} | \psi_{\alpha}}}.
\label{phase rigidity}
\end{equation}
Hermitian systems have $r_{\alpha}=1$ for all $\alpha$, but when approaching an EP, $r_{\alpha}\rightarrow 0$ for the states that coalesce.

Exceptional points lead to intriguing phenomena such as unidirectional invisibility~\cite{Lin2011}, single-mode lasers~\cite{Feng2014,Hodaei2014}, or enhanced sensitivity in optics~\cite{Chen2017,Hodaei2017,Rechtsman2017}. Many of these phenomena could be understood in terms of an environment mediated interaction~\cite{alvarez2006,Bird2004,Yoon2012,Rotter2015,Auerbach2011,Eleuch_PhysRevA_2017}.

\section{Challenges for establishing a bulk-boundary correspondence in non-Hermitian lattices}
\label{sec:bulk-boundary}

In this section we examine a few key challenges for non-Hermitian lattices. To start with we present a simple one-dimensional model that will serve to fix ideas.
\subsection{Simple ladder model}

\begin{figure}
\centering
\includegraphics[width=0.7\linewidth]{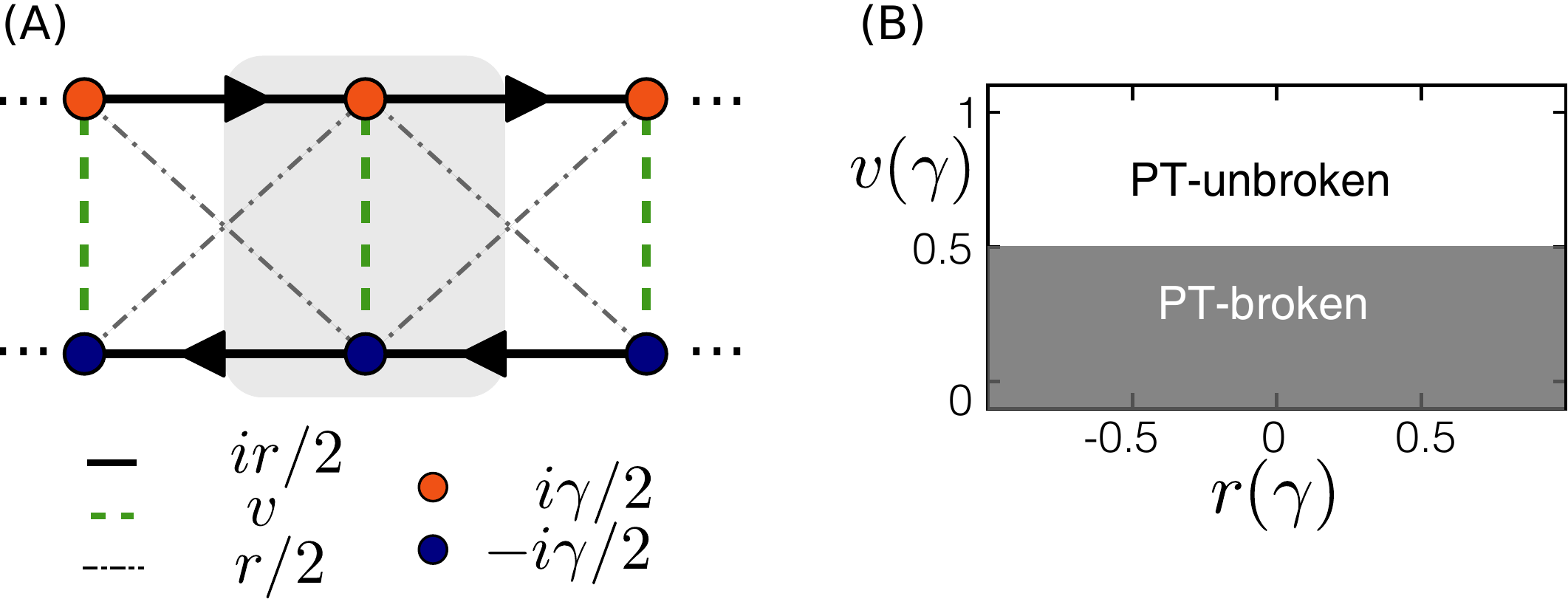}
\caption{(A) Scheme representing three cells of the non-Hermitian lattice model~\cite{lee_anomalous_2016} introduced in the text. (B) Map of the PT-broken and unbroken regions as a function of the Hamiltonian parameters $v$ and $r$ (both in units of $\gamma$).}
\label{Fig_1}
\end{figure}

To motivate our discussion let us consider the one-dimensional non-Hermitian model of Ref.~\cite{lee_anomalous_2016}:
\begin{equation}
{\cal H}_k=h_x(k)\sigma_x + \left(h_z(k) + \frac{i\gamma}{2}\right)\sigma_z,
\label{Hamiltonian}
\end{equation}
where $\sigma_x,\sigma_z$ are the Pauli matrices, $\gamma$ is a real parameter controlling the degree of non-Hermiticity of ${\cal H}_k$ and $k$ is the wave-vector.  $h_x(k)=v+r\cos k$ and  $h_z(k)=r\sin k$ are chosen as to encircle the EP located at $(h_x,h_z)=(\pm \gamma/2,0)$ ($v$ and $r$ are real parameters). The model can be represented by a tight-binding lattice with gains and losses on different sublattices as represented in Fig.~\ref{Fig_1}(A). This model can be realized in an array of coupled
resonator optical waveguides~\cite{Hafezi2011} and in a photonic crystal~\cite{Weimann2017} as pointed out by Lee~\cite{lee_anomalous_2016}. 

The Hamiltonian of Eq.~(\ref{Hamiltonian}) commutes with the composed parity-time, ${\cal PT}$, operator~\cite{lee_anomalous_2016}. PT-symmetry is said to be unbroken if all Hamiltonian eigenstates are also eigenstates of the ${\cal PT}$ operator, thereby having real-values eigenenergies (see Fig.~\ref{Fig_1}(B)).
Curiously, this model has a \textit{single} zero-energy edge state localized on one side of the system (and not one on each side). This kind of behavior seems at odds with the bulk-boundary correspondence of Hermitian systems and motivated the definition of a fractional winding number of $1/2$ in Ref.~\cite{lee_anomalous_2016} and other related works~\cite{leykam_edge_2017,Shen_topological_2017,yin2018geometrical}. Dynamical phase transitions were also addressed for this model in Ref.~\cite{zhou_dynamical_2017}. We will discuss additional properties of this Hamiltonian in the next subsection.

Another paradigmatic Hamiltonian is the non-Hermitian version of the Su-Schrieffer-Heeger model: ${\cal H}_k=h_x(k)\sigma_x + \left(h_y(k) + \frac{i\gamma}{2}\right)\sigma_y$ and its quasi-one-dimensional variants~\cite{Ryu_Antiresonance_2017,ryu_reconfiguration_2017}. These have been studied in several recent papers by Yuce~\cite{Yuce_Edge_2018}, 
Yao and Wang~\cite{yao_edge_2018}, Lieu~\cite{Lieu_2018} and Yin \textit{et al.}~\cite{yin2018geometrical}. Other studies went even further to study the interplay between non-Hermiticity and driving in this type of simple models~\cite{yuce_topological_2015}, or others with non-reciprocal interactions~\cite{li2018floquet}\footnote{There has been much interest in bringing non-Hermiticity to Floquet systems to achieve non-reciprocal effects~\cite{koutserimpas2018nonreciprocal,longhi2017non}. Other related work uses a bipartite system instead of non-Hermiticity~\cite{dal_lago_one-way_2017}.}.
 
\subsection{Challenges for a bulk-boundary correspondence}
\label{challenges}

In the context of topological insulators, usually, when one refers to a state being ``topological" it is meant that there is an associated \textit{bulk-boundary correspondence}~\cite{bernevig_topological_2013,frank_ortmann_topological_2015,Asboth_Short_2016}. The bulk-boundary correspondence is one of the milestones behind the theory of topological states of matter and represents a relation between a bulk property of a (translational-invariant) material or lattice encoded in a topological invariant (obtained from the Bloch-type eigenstates) and what happens at its boundary (surface, edge, etc). Typically, this invariant can predict the number of boundary states and chirality. Interestingly, the states which are bound to appear at the boundary are robust to imperfections and defects, after all their existence is tied to a non-local quantity (the topological invariant), arising from integration over the Brillouin zone of a kernel which depends on the bulk eigenstates. Examples of topological invariants include the Chern number and the Zak phase~\cite{bernevig_topological_2013,Asboth_Short_2016}.   

\begin{figure}
\centering
\includegraphics[width=0.7\linewidth]{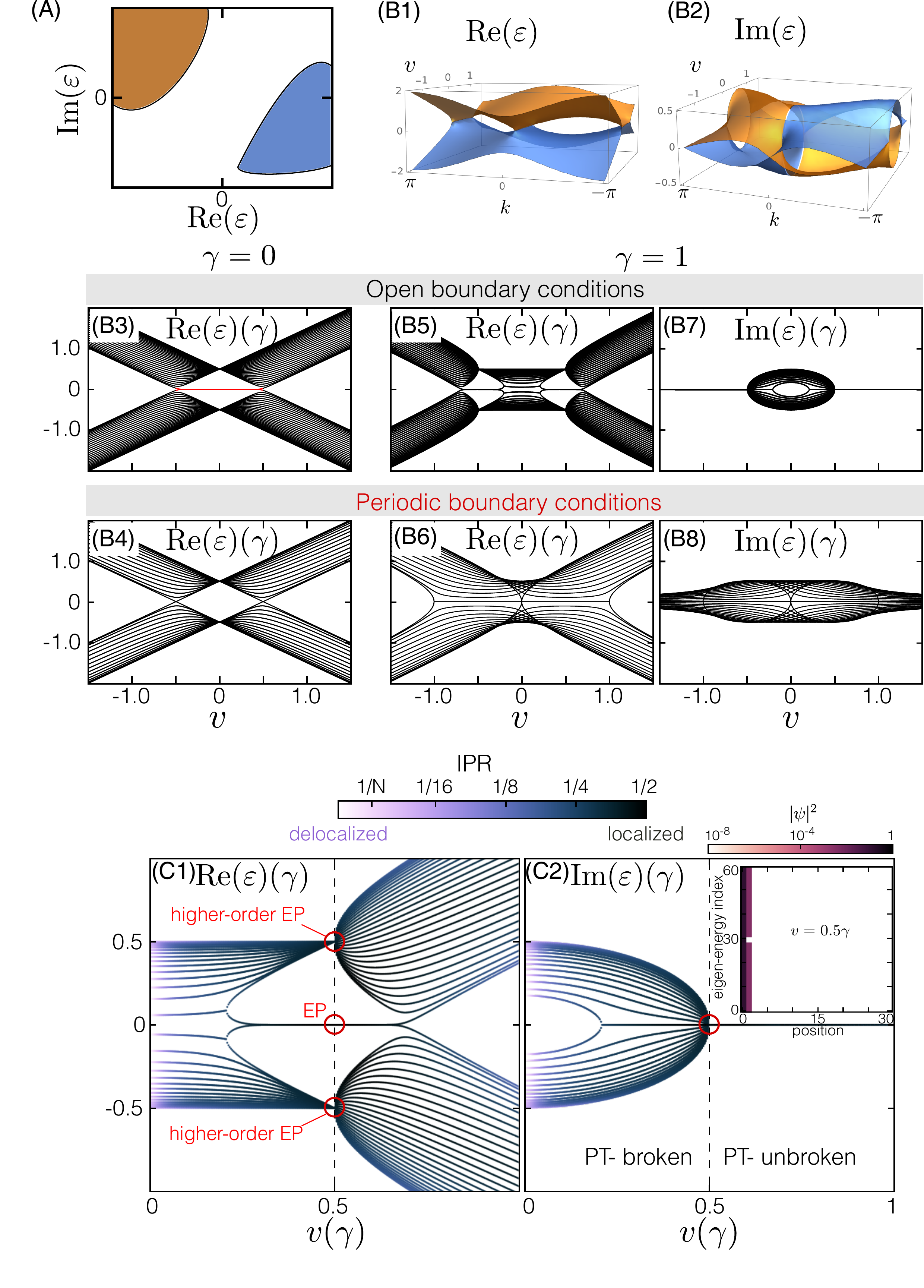}
\caption{(A) shows two generic bulk bands (in gold and blue) in the complex plane. The bulk bands are isolated according to the definition in Ref.~\cite{Shen_topological_2017}. (B1) and (B2) show the dispersion relation versus $k$ and $v$ for the model of Eq.(\ref{Hamiltonian}) with $N=30$, $v=r=0.5$ in the translational invariant case. Panels (B3-B8) show the real and imaginary parts of the eigenenergies for open (B3, B5 and B7) and periodic boundary conditions (B4, B6 and B8) obtained for the model of Eq.(\ref{Hamiltonian}) as a function of the parameter $v$ (with $N=30$, $r=0.5$). The left most panels (B3 and B4) are for the Hermitian case, $\gamma=0$ (here we do not show the imaginary part of the spectrum as it vanishes), while the others have $\gamma=1$. The edge states in the Hermitian case are colored in red. (C1) Real and (C2) imaginary part of the eigen-energies obtained as a function of $v$. This corresponds to a finite system with $N=30$ and $r=0.5$ ($\gamma$ is taken here as the unit of energy). The color scale shows the inverse participation ratio. The inset in (C2) shows the probability density associated to the eigenstates obtained for $v=0.5$. In contrast with the Hermitian case (B3 and B4), one must notice that all states remain localized at one edge as evidenced by the color scale. Panels (C1)  and (C2) are adapted from Ref.~\cite{Martinez_Non-Hermitian_2018} with permission.}
\label{Fig_2}
\end{figure}

Therefore, a very first step in the build-up of a theory for topological states in non-Hermitian lattices relies on establishing a bulk-boundary correspondence. Here we find a first crucial challenge. Indeed, the existence of such correspondence is at the center of an intense debate~\cite{hu_absence_2011,esaki_edge_2011,lee_anomalous_2016,Xiong_Why_2018,Shen_topological_2017}, a discussion which has also been inspired by the model introduced in the previous section~\cite{lee_anomalous_2016,Xiong_Why_2018,Martinez_Non-Hermitian_2018}. There are three uncomfortable facts of non-Hermitian lattices that are challenging in this context: 

\begin{enumerate}[(A)]
\item \textbf{Defining a gap for a complex spectrum.} In contrast with real numbers, there is no order relation for complex numbers. Therefore, a first obstacle is how to define a gap among energy bands with both real and imaginary parts. Let us consider a non-Hermitian lattice (a periodic system) with eigenstates of the Bloch form and whose energies are $E_n(\bk)$, where $ \bk $ is the crystal momentum in the Brillouin zone and $n$ the band index. One of the proposals to generalize the notion of gapped system is that of Ref.~\cite{Shen_topological_2017} where the authors define a band $ n $ to be ``separable'' if its energy $E_n(\bk) \neq E_m(\bk)$ for all $m \neq n$ and all $\bk$;  ``isolated'' if $E_n(\bk) \neq E_m(\bk')$ for all $m \neq n$ and all $\bk, \bk'$ (see the scheme in Fig.~\ref{Fig_2}(A)) \footnote{This is, the region $\{ E_n(\bk), \bk \in {\rm BZ} \}$ (now in the complex plane) does not overlap with that of any other band. In this situation, the authors say that ``the band $E_n(\bk)$ is surrounded by a ``gap'' in the complex energy plane where no bulk states exist"~\cite{Shen_topological_2017}.}; ``inseparable'' when for some momentum the (complex) energy becomes degenerate with another band. Other authors take a different approach, generalizing the concept of bandgap
as the prohibition of touching a (generally complex) base energy~\cite{gong2018topological}.

\item \textbf{Extreme sensitivity of the spectrum to boundary conditions.} A second difficulty is that non-Hermitian systems tend to be extremely sensitive to boundary conditions, and therefore the expected correspondence between the eigen-energies of a (sufficiently large) system with open boundary conditions and those obtained for periodic boundary conditions does not hold~\cite{lee_anomalous_2016,Xiong_Why_2018}. This difficulty was emphasized by Xiong~\cite{Xiong_Why_2018} and one can appreciate it already for the simple model of the previous section as shown in Fig.~\ref{Fig_2}(B)). In this case, even in the large system limit the spectra for the two boundary conditions do not match, see Figs.~\ref{Fig_2}(B5-B8) (the real and imaginary parts of the energy for the infinite case as a function of the wavevector $k$ and the parameter $v$ is shown in Figs.~\ref{Fig_2}(B1-B2)), which contrasts with the Hermitian case (Figs.~\ref{Fig_2}(B3-B4)) where except for the boundary states (shown in red) the spectrum matches.

\item \textbf{An ordered non-Hermitian lattice may have all eigenstates (anomalously) localized at the boundary.} Closely related to the previous point, for non-Hermitian lattices one may have a situation where a finite but arbitrarily large system does not contain any extended (Bloch-like) states and all states are localized at the boundary~\cite{Xiong_Why_2018,Martinez_Non-Hermitian_2018}, even those that are supposed to be deep in the bands. This poses a conceptual problem even deeper than point (B), even if we get a way to make bulk and boundary to match from the viewpoint of eigenvalues, the character of the states seems unbridgeable.

This is the case for example for the model of Ref.~\cite{lee_anomalous_2016} introduced in the previous subsection. In Ref.~\cite{Martinez_Non-Hermitian_2018} this anomalous localization (which occurs even when the system has no disorder) was attributed to exceptional points of higher order, where a macroscopic fraction of the states of an (eventually large) system coalesce, thereby devoiding the system of extended states.

Figure~\ref{Fig_2}(C) shows these higher order EPs. The localization of the full bandstructure is evidenced in the color scale showing the inverse participation ratio. The \textit{inverse participation ratio} (IPR) is a measure of localization of a state $\psi_{\alpha}$~\cite{Kramer_1993,Evers2008}:
\begin{equation}
I_{\alpha}= \sum_{\textbf{r}} |\psi_{\alpha}(\textbf{r})|^4 / \left(\sum_{\textbf{r}} |\psi_{\alpha}(\textbf{r})|^2 \right)^2.
\label{ipr}
\end{equation}
The inverse of this number is essentially the average diameter of the state (in one-dimension). For extended states, $1/I_{\alpha}$ is the system's volume $L^d$. In Fig.~\ref{Fig_2}(B5-B8) we can see that close to $v=0.5\gamma$ all eigenstates, even those in the bands, are localized. Here, must notice the stark contrast with the Hermitian case where given a gapped Hamiltonian (as occurs for some values of $v$ in Fig.~\ref{Fig_2}(B3)(B4)) one would immediately be drawn to the states in the gap (marked in red) which are the natural boundary state candidates. Here, all the available states are localized (as shown by the color scale in Fig.~\ref{Fig_2}(B5-B8)) over a broad range of $v$ values.

This peculiar type of localization (different from other mechanisms such as Anderson localization) of all eigenstates at one edge was shown to be robust to moderate amounts of disorder, even when the origin of this robustness is not topological in the sense of any known underlying bulk-boundary correspondence.
Quantities such as the phase-rigidity also show consequences which pervade even far from these higher-order exceptional points~\cite{Martinez_Non-Hermitian_2018}. The mechanism was tied to an environment mediated interaction effect.

The same effect was found analytically in Ref.~\cite{yao_edge_2018} for a non-Hermitian SSH model and dubbed \textit{non-Hermitian skin effect}.

\end{enumerate}

\section{Proposals for a classification of topological states in non-Hermitian lattices}
\label{sec:classification}

The literature devoted to finding topological states in non-Hermitian lattices is growing at rapid pace. Classifying these states and the possible topological phases becomes crucial in this first stage of this emerging research front. One of such proposals aiming at a systematic and consistent classification was given in Ref.~\cite{Shen_topological_2017}. The definitions of ``separable'', ``isolated'' and ``inseparable'' bands introduced in~\cite{Shen_topological_2017} and mentioned in Section \ref{challenges}
aim to generalize the ideas of gapped, fully gapped and gapless bands in the Hermitian case, and form the basis of the topological classification presented in Ref.~\cite{Shen_topological_2017}. The authors present a generalization of the Chern number in two-dimensions together with a new classification for the one-dimensional case. Interestingly, in the latter case and unlike for Hermitian systems, the topology is determined by the energy dispersion rather than the energy eigenstates~\cite{Shen_topological_2017}, the \textit{vorticity} of the eigenvalues. One must notice, however, that one of the assumptions is that the bandstructure is separable, thereby ruling out cases where the degeneracies may introduce interesting effects.

Another interesting proposal to bring a unified classification is that of Ref.~\cite{gong2018topological} which is based on two main assumptions:
Topological phases of non-Hermitian systems can be understood as dynamical phases where the imaginary part of the eigenvalues is relevant; and the generalization of the concept
of the band gap is taken by the prohibition of touching a (generally complex) base energy.
The authors find a classification in analogy with the Hermitian periodic table in the Altland Zirnbauer (AZ) classes in all dimensions. One of the main conclusions is the absence of non-Hermitian topological phases in two dimensions~\cite{gong2018topological}. This seems at odds with the conclusions of Ref.~\cite{Shen_topological_2017} and ~\cite{leykam_edge_2017}. The authors also emphasize that exceptional points ``while unique to non-Hermitian systems and  of great experimental interest, may not be a good starting point for a systematic classification, since they imply band touching in the bulk and seem incompatible with a non-Hermitian generalization of gap''~\cite{gong2018topological}. This marks a clear departure point from the other classification attempts. Although pragmatic, leaving exceptional points outside of the game may seem an expensive price in a field which was largely established around them as objects of study. 

Finally, Ref.~\cite{rudner_survival_2016} focused on lattices in one dimension with losses added as on-site terms identifying a winding number, and other authors where moved to choose non-Bloch invariants~\cite{yao_non-hermitian_2018} in an attempt to bridge the challenges to establish a bulk-boundary correspondence.

\section{Experimental prospects}
\label{sec:experiments}

Over the last few years there has been a plethora of experiments elucidating different aspects or new phenomenology associated to non-Hermitian systems~\cite{Rotter2015} including: 
the observation of exceptional points and their peculiar properties~\cite{dembowski2001,stehmann2004,hahn2016observation}, 
non-reciprocal effects in optics~\cite{nazari2014optical}, the observation of
topological edge states in PT -symmetric quantum walks~\cite{xiao2017observation},
topological energy transfer in optomechanical systems with exceptional points~\cite{xu2016topological,xu_topologicaldyn_2017} and topological plasmons~\cite{downing2018topological}.

In the context of topological transitions, inspired by earlier predictions~\cite{rudner_topological_2009,diehl_topology_2011,esaki_edge_2011,bardyn2013,schomerus2013topologically,yuce_topological_2015}, experiments using optical waveguide arrays demonstrated a scheme to extract winding number of one-dimensional systems from bulk dynamics measurements~\cite{zeuner2015observation}.

The selective amplification of topologically protected localized midgap states (in systems with spatially distributed gain and loss)
 proposed in Ref.~\cite{schomerus2013topologically} has been demonstrated in microwave experiments~\cite{poli2015selective}. Furthermore, it was shown that topological states absent when the system is Hermitian can be induced by adding losses~\cite{Malzarḍ_2015}.
 
The possibility of realizing a synthetic magnetic flux in photonic lattices either through balanced gain and loss~\cite{leykam2017flat} or periodic modulations~\cite{fang_realizing_2012,Longhi_2014,mukherjee_experimental_2018}
opens fascinating prospects including the observation of the photonic analogue of Aharonov-Bohm caging (an effect where for certain geometries at specific values of a perpendicular magnetic field, non-interacting particles become fully localized)~\cite{fang_realizing_2012,mukherjee_experimental_2018}.

\begin{figure}
\centering
\includegraphics[width=0.7\linewidth]{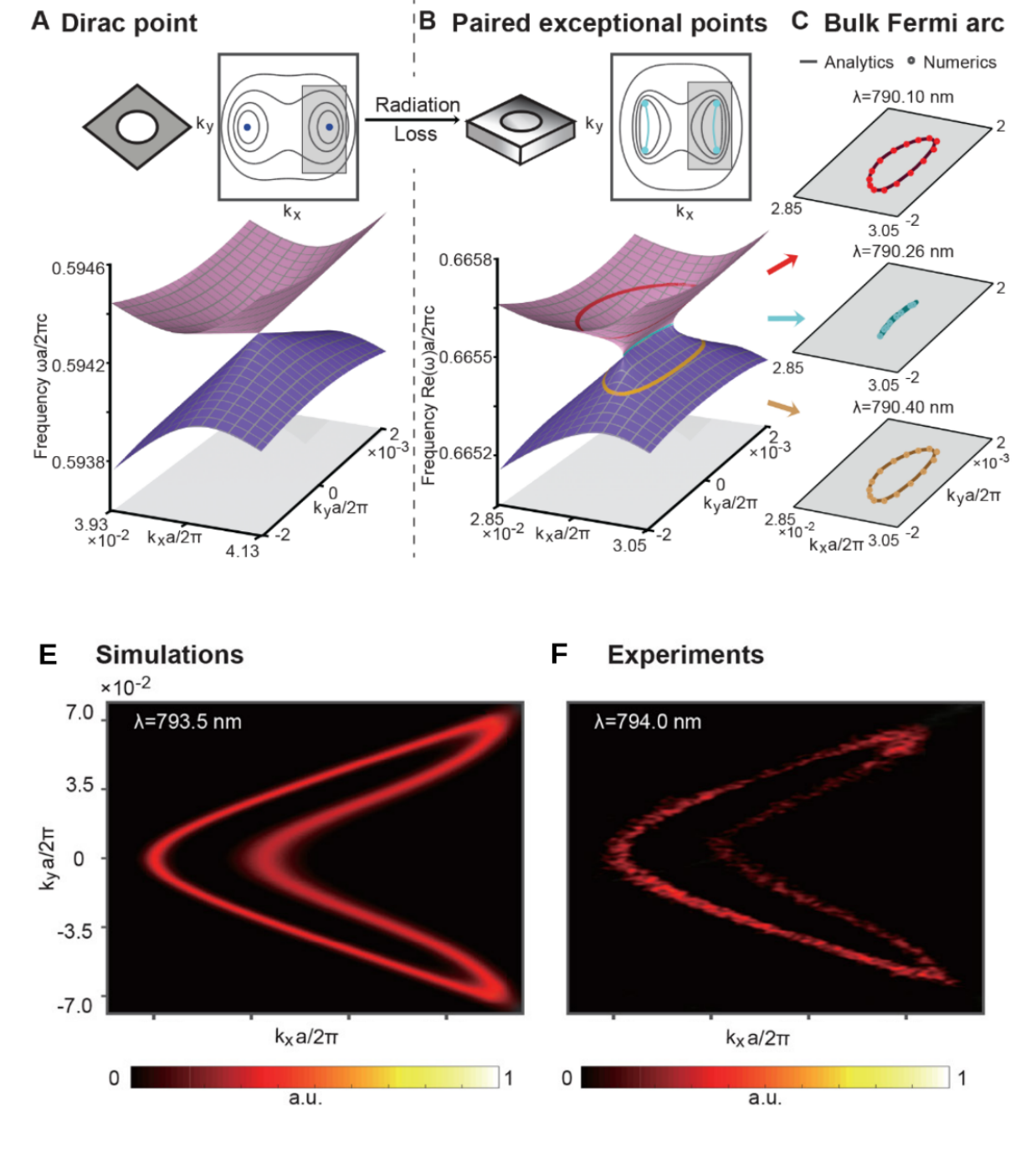}
\caption{Illustration a bulk Fermi arc arising when a Dirac point split into paired exceptional points~\cite{zhou2018observation}. (A) and (B) show the associated photonic crystal structures, isofrequency contours, and band structures. The banstructure in (A) is for a 2D rhombic lattice of elliptical air holes and has a single Dirac point as shown. (B) shows the real part of the eigenenergies of an open system formed by a 2D slab with finite thickness. The real part of the eigenenergies are degenerate along the open-ended contour which form a Fermi arc connecting the pair of exceptional points. (C) shows examples of isofrequency contours. Solid lines are from the analytical model, and circles are from numerical simulations. (E) and (F) show the simulated and experimental isofrequency contours for the indicated wavelengths. Figures from [Zhou \textit{et al.} \href{http://science.sciencemag.org/content/early/2018/01/10/science.aap9859.full}{Science eaap9859 (2018)}]. Reprinted with permission from AAAS.}
\label{Fig_3}
\end{figure}

Finally, among the unconventional phenomena coming from the subtle interplay between non-Hermiticity and topology we would like to mention the tachyon-like dispersions~\cite{szameit2011p} demonstrated in photonic crystals~\cite{zhen2015spawning}
and the experimental observation of bulk Fermi arcs arising from radiative losses in an open system of photonic crystal slabs~\cite{zhou2018observation} illustrated in Fig.~\ref{Fig_3}.

\section{Final remarks}
\label{sec:conclusions}
Although the general principles emerging from the field of  topological insulators are today much celebrated, the need of crosstalk among communities is still of utmost importance. This may be particularly useful in the search topological states in non-Hermitian systems where there is a large potential for synergy between the communities of topological insulators, photonics, mathematical physics, quantum physics and ultracold matter.

In this brief overview we have addressed a small part of the rich variety of new phenomena brought by the non-Hermiticity in lattice systems. The prospects are fascinating as most of the paradigms taken from granted in the theory of topological states of matter, e.g. the existence of Bloch-type states deep in the bands and the very definition of energy gap, are challenged. This is reflected in the rapidly growing literature and the many open discussions, including the contrasting proposals for a classification. Everything seems to indicate that we are just entering a promising \textit{terra incognita}.

\begin{acknowledgements}
LEFFT, JEBV and MB acknowledge support from FondeCyT (Chile) under grants number 1170917, 3170126, and 3170143. 

\end{acknowledgements}


%\begin{thebibliography}{}
% etc
%\end{thebibliography}

\end{document}